\documentclass[12pt]{article}

\ifx\pdfoutput\undefined
\usepackage[dvips,bookmarks]{hyperref}
\else
\usepackage{hyperref}
\fi
\hypersetup{colorlinks=false,bookmarksopen,bookmarksnumbered,citecolor=blue,
   pdfstartview=FitH}

\usepackage{latexsym}
\usepackage{amssymb,amsfonts,amsmath}
\usepackage{indentfirst}
\usepackage{bbm}

\parskip=\medskipamount

\newcommand {\cD}{{\cal D}}
\newcommand {\cE}{{\cal E}}

\newcommand {\cK}{{\cal K}}
\newcommand {\cL}{{\cal L}}
\newcommand {\cM}{{\cal M}}
\newcommand {\cN}{{\cal N}}

\newcommand {\cT}{{\cal T}}
\newcommand {\cU}{{\cal U}}

\def\a{\alpha}

\def\b{\beta}

\def\d{\delta}

\def\g{\gamma}

\def\q{\theta}
\def\r{\rho}
\def\s{\sigma}
\def\t{\tau}

\def\F{\Phi}

\def\O{\Omega}

\def\S{\Sigma}

\newcommand {\rd}{{\rm d}}
\newcommand {\ri}{{\rm i}}
\newcommand {\re}{{\rm e}}

\newcommand{\ad}{{\dot{\alpha}}}                           %new
\newcommand{\bd}{{\dot{\beta}}}                            %new
\newcommand{\ve}{\varepsilon}                            %new
\newcommand{\cDB}{{\bar\cD}}                            %new

\newcommand{\pa}{\partial}                           %new
\newcommand{\hf}{\frac12}
\newcommand{\be}{\begin{equation}}
\newcommand{\ee}{\end{equation}}
\newcommand{\bea}{\begin{eqnarray}}
\newcommand{\eea}{\end{eqnarray}}
\newcommand{\non}{\nonumber}
\newcommand{\1}{\underline{1}}
\newcommand{\2}{\underline{2}}

\newcommand{\bm}[1]{\mbox{\boldmath$#1$}}

\def\double #1{#1{\hbox{\kern-2pt $#1$}}}

\newcommand{\gd}{{\dot\g}}
\newcommand{\dd}{{\dot\d}}

\newcommand{\ts}{{\tilde{\s}}}

\newcommand{\alu}{{\underline{\a}}}
\newcommand{\beu}{{\underline{\b}}}
\newcommand{\gau}{{\underline{\g}}}

\newcommand{\sba}{{\bar{\s}}}

\newcommand{\teb}{{\bar{\theta}}}

\begin{document}
%%%%%%%%%%%%%%%%
%%%%%%%%%%%%%%%%
\begin{titlepage}
\begin{flushright}
UUITP-13/09\\
YITP-SB-09-09\\
UMD-PP-09-035\\
August, 2009\\
\end{flushright}
\vspace{5mm}

\begin{center}
{\Large \bf On conformal  supergravity and projective superspace}\\ 
\end{center}

\begin{center}

{\bf
S. M. Kuzenko\footnote{kuzenko@cyllene.uwa.edu.au}${}^{a}$,
U. Lindstr\"om\footnote{ulf.lindstrom@teorfys.uu.se}${}^{b}$,
M. Ro\v{c}ek\footnote{rocek@max2.physics.sunysb.edu}${}^{c}$, 
G. Tartaglino-Mazzucchelli\footnote{gtm@umd.edu}${}^{a,d}$
} \\
\vspace{5mm}

\footnotesize{
${}^{a}${\it School of Physics M013, The University of Western Australia\\
35 Stirling Highway, Crawley W.A. 6009, Australia}}  
~\\
\vspace{2mm}

\footnotesize{
${}^{b}${\it Department of Theoretical Physics,
Uppsala University \\ 
Box 803, SE-751 08 Uppsala, Sweden}
}
\\
\vspace{2mm}

${}^c${\it C.N.Yang Institute for Theoretical Physics, Stony Brook University \\
Stony Brook, NY 11794-3840,USA}\\
\vspace{2mm}

${}^{d}${\it Center for String and Particle Theory,
Department of Physics, 
University of Maryland\\
College Park, MD 20742-4111, USA}
~\\

\end{center}
\vspace{5mm}

\begin{abstract}
\baselineskip=14pt
The projective superspace formulation for four-dimensional  $\cN=2$ matter-coupled supergravity
presented in arXiv:0805.4683 makes use of the variant superspace realization for 
the $\cN=2$ Weyl multiplet in which the structure group 
is ${\rm SL}(2,{\mathbb C})\times {\rm SU}(2)$ and  the super-Weyl transformations 
are generated by a covariantly chiral parameter. An extension to 
Howe's realization of $\cN=2$ conformal supergravity in which the tangent space  group 
is ${\rm SL}(2,{\mathbb C})\times {\rm U}(2)$ and  the super-Weyl transformations 
are generated by a real unconstrained parameter
was briefly sketched. Here we give the explicit details of the extension.
\end{abstract}
\vspace{1cm}

\vfill
\end{titlepage}

\newpage
\renewcommand{\thefootnote}{\arabic{footnote}}
\setcounter{footnote}{0}

%%%%%%%%%%%%%%%%%%%%%%%%%%%%%%%%%%%%%%%%%%%%%%%%%%
%%%%%%%%%%%%%%%%%%%%%%%%%%%%%%%%%%%%%%%%%%%%%%%%%%

\section{Introduction}
\setcounter{equation}{0}
Long ago, Howe \cite{Howe}
proposed superspace formulations for four-dimensional 
$\cN\leq 4$ conformal supergravity theories
\cite{KTvN,FZ,Bergshoeff:1980sw,Bergshoeff:1980is}
by explicitly gauging ${\rm SL}(2,{\mathbb C})\times {\rm U}(\cN) $ 
and identifying appropriate constraints on the torsion of curved superspace.
In the case $\cN=1$, which had been earlier  elaborated in a somewhat different 
but equivalent setting in \cite{HT},  
the approach of \cite{Howe} was utilized \cite{GGRS} 
to provide a unified description for  the known off-shell realizations 
({\it i.e.}, the old minimal, new minimal and non-minimal formulations) 
for $\cN=1$ Poincar\'e supergravity and the corresponding matter couplings.
In the $\cN=2$ case, few applications of Howe's formulation
have appeared -- essentially only the  demonstration in \cite{Howe,Muller} of
how to obtain some off-shell formulations for pure $\cN=2$ Poincar\'e
supergravity by coupling the Weyl multiplet to compensating multiplets, 
generalizing the $\cN=2$ superconformal tensor calculus \cite{tensor}.
No general discussion of matter couplings within the superspace 
setting of \cite{Howe} has been given.
Of course, there is a simple historical explanation for that. 
Even in rigid $\cN=2$ supersymmetry, the adequate approaches 
for generating off-shell supermultiplets and supersymmetric actions 
appeared only in 1984; they go under the names {harmonic superspace}
\cite{GIKOS,GIOS} and {projective superspace} \cite{KLR,GHR,LR,LR2}.\footnote{The 
relationship between the {\it rigid} harmonic and projective superspace formulations
is spelled out in \cite{K_double}. For a recent discussion, see also \cite{Jain:2009aj}.} 
The relation of the approach of \cite{Howe} to the harmonic superspace 
formulation for $\cN=2$ supergravity and its matter couplings 
\cite{Galperin:1987ek,GIOS}  has not been elucidated in detail, 
except for a short and incomplete discussion in \cite{HH}.

A year ago, we developed a projective superspace formulation for
4D ${\cal N}=2$ supergravity and its matter couplings \cite{KLRT_M}.\footnote{The 
harmonic and projective superspace approaches to 
$\cN=2$ matter-coupled supergravity differ in 
(i) the structure of covariant off-shell   supermultiplets used; 
and (ii) the locally supersymmetric action principle chosen. }
In that work, we used an alternative superspace formulation for ${\cN=2}$ conformal 
supergravity.
It differs from that given in  \cite{Howe} in the following three points: 
(i) the structure group is identified with 
${\rm SL}(2,{\mathbb C})  \times  {\rm SU(2)}$; (ii) the geometry of curved superspace is
 subject to the constraints introduced by Grimm \cite{Grimm}; 
 (iii)  the  super-Weyl transformations are generated 
 by a covariantly chiral but otherwise unconstrained superfield.
 In \cite{KLRT_M}, we also briefly sketched the correspondence between 
 the two superspace formulations for conformal supergravity.
 In the present note, we explicitly extend the approach of \cite{KLRT_M} to the case of Howe's 
formulation for conformal supergravity.

This paper is organized as follows. 
In section 2 we first review the formulation of \cite{Howe}
for $\cN=2$ conformal supergravity, and present the finite form 
for the corresponding super-Weyl transformations.
Using the latter result, we demonstrate how the  formulation 
used in \cite{KLRT_M} emerges from Howe's formulation upon  
gauge fixing the super-Weyl and local U(1) symmetries. 
In section 3 we introduce a family of covariant projective supermultiplets 
and propose a locally supersymmetric and super-Weyl invariant action principle. 

\section{Conformal supergravity}
\setcounter{equation}{0}
We start by reviewing the superspace formulation for $\cN=2$ conformal supergravity 
proposed in \cite{Howe}.

\subsection{Superspace geometry of conformal supergravity}
\label{subsec3.1}
Consider a curved four-dimensional  $\cN=2$ superspace  $\cM^{4|8}$ parametrized by local 
coordinates  $z^{{M}}=(x^{m},\q^{\mu}_\imath,{\bar \q}_{\dot{\mu}}^\imath)$,
where $m=0,1,\cdots,3$, $\mu=1,2$, $\dot{\mu}=1,2$ and  $\imath= \1,\2$.
The Grassmann variables $\q^{\mu}_\imath $ and $\teb_{\dot{\mu}}^\imath$
are related to each other by complex conjugation: 
$\overline{\q^{\mu}_\imath}=\teb^{\dot{\mu}\imath}$. 
Following \cite{Howe}, we choose the structure group 
to be ${\rm SL}(2,{\mathbb C})\times {\rm SU}(2)_R \times {\rm U}(1)_R$,
and let $M_{ab}=-M_{ba}$, $J_{ij}=J_{ji}$ and $\mathbb J$ be the corresponding 
Lorentz, ${\rm SU}(2)_R$ and ${\rm U}(1)_R$ generators.
Along with  gauge fields for the three subgroups of the structure group,
which are necessary to describe the multiplet of conformal supergravity,   
it is also useful to introduce an Abelian vector multiplet associated with 
 an internal group ${\rm U}(1)_Z$ with generator $\mathbb Z$ such that 
$[M_{ab} , {\mathbb Z} ] = [J_{ij} , {\mathbb Z} ] =[{\mathbb J} , {\mathbb Z} ] =0$.
One can think of  $\mathbb Z$ as a central charge operator.
The central charge vector multiplet contains the graviphoton.
The covariant derivatives 
$\cD_{{A}} =(\cD_{{a}}, \cD_{{\a}}^i,\cDB^\ad_i) 
\equiv (\cD_{{a}}, \cD_{ \underline{\a} }, \cDB^{\underline{\ad}})$ 
have the form 
\bea
\cD_{A}&=&E_{A}+
\hf \,\O_{A}{}^{bc}\,M_{bc}+
\Phi_{A}^{~\,kl}\,J_{kl}
+ \ri \,\Phi_{A}\,{\mathbb J}
+V_A\, {\mathbb Z} \non\\
&=&E_{A}~+~\O_{A}{}^{\b\g}\,M_{\b\g}
+{\O}_{A}{}^{\bd\gd}\,\bar{M}_{\bd\gd}
+\Phi^{~\,kl}_{A}\,J_{kl}
+\ri \, \Phi_{A}\,{\mathbb J}
+V_A\, {\mathbb Z}~.
\label{CovDev}
\eea
Here $E_{{A}}= E_{{A}}{}^{{M}} \pa_{{M}}$ is the supervielbein, 
with $\pa_{{M}}= \pa/ \pa z^{{M}}$,
$\O_{{A}}{}^{bc}$ is the Lorentz connection, 
 $\Phi_{{A}}{}^{kl}$ and $\Phi_{{A}}$ are  
 the ${\rm SU}(2)_R$ and ${\rm U}(1)_R$ connections, respectively.
Finally, the vector multiplet is described by  $V_A$.

The Lorentz generators with vector indices ($M_{ab}$) and spinor indices
($M_{\a\b}=M_{\b\a}$ and ${\bar M}_{\ad\bd}={\bar M}_{\bd\ad}$) are related to each other 
by the standard rule:
$$
M_{ab}=(\s_{ab})^{\a\b}M_{\a\b}-(\tilde{\s}_{ab})^{\ad\bd}\bar{M}_{\ad\bd}~,~~~
M_{\a\b}=\hf(\s^{ab})_{\a\b}M_{ab}~,~~~
\bar{M}_{\ad\bd}=-\hf(\tilde{\s}^{ab})_{\ad\bd}M_{ab}~.
$$ 
The generators of the structure group
act on the spinor covariant derivatives as follows:\footnote{The 
(anti)symmetrization of $n$ indices 
is defined to include a factor of $(n!)^{-1}$.}
\bea
{[}M_{\a\b},\cD_{\g}^i{]}
&=&\ve_{\g(\a}\cD^i_{\b)}~,\qquad
{[}\bar{M}_{\ad\bd},\cDB_{\gd}^i{]}=\ve_{\gd(\ad}\cDB^i_{\bd)}~, \non \\
{[}J_{kl},\cD_{\a}^i{]}
&=&-\d^i_{(k}\cD_{\a l)}~,
\qquad
{[}J_{kl},\cDB^{\ad}_i{]}
=-\ve_{i(k}\cDB^\ad_{l)}~, \non \\
{[}{\mathbb J},\cD_{\a}^i{]} &=&\cD_{\a}^i~,\qquad  \qquad ~~\,
{[}{\mathbb J},\cDB^{\ad}_i{]}~
=\,-\cDB^{\ad}_i~,
\label{generators}
\eea
while $ [{\mathbb Z},\cD_A]=0$.
Our notation and conventions coincide with those adopted in \cite{KLRT_M} 
and correspond to \cite{BK}.

The entire gauge group is generated by local transformations
of the form 
\be
\d_\cK \cD_{{A}} = [\cK  , \cD_{A}]~,
\qquad \cK = K^{{C}} \cD_{{C}} +\hf K^{ c  d} M_{c  d}
+K^{kl} J_{kl}  +\ri L\, {\mathbb J} + \t \,{\mathbb Z}~,
\label{tau}
\ee
with the gauge parameters
obeying natural reality conditions, but otherwise  arbitrary. 
Given a tensor superfield $\cU(z)$, with its indices suppressed, 
it transforms as follows:
\bea
\d_\cK \cU = \cK\, \cU~.
\label{tensor-K}
\eea

The  covariant derivatives obey the algebra
\bea
{[}\cD_{{A}},\cD_{{B}}\}&=&T_{{A}{B}}{}^{{C}}\cD_{{C}}
+\hf R_{{A}{B}}{}^{{c}{d}}M_{{c}{d}}
+R_{{A}{B}}{}^{kl}J_{kl}
+\ri R_{{A}{B}}\, {\mathbb J}
+F_{AB}\, {\mathbb Z}
~,
\label{algebra}
\eea
where $T_{{A}{B}}{}^{{C}}$ is the torsion, $R_{{A}{B}}{}^{kl}$, $R_{{A}{B}}$ 
and $R_{{A}{B}}{}^{{c}{d}}$ are the curvatures and 
$F_{AB}$ the vector multiplet field strength.
To describe conformal supergravity, the torsion has to be subject to
the following constraints \cite{Howe}: 
\bea
T_{\alu\beu}{}^{C}&=& T_{\alu}{}^{\dot{\beu}\, \gau}=0~, 
\qquad 
T_{\a}^i{}^\bd_j{}^c=-2\ri\d^i_j(\s^{c})_{\a}{}^{\bd}~, \non \\
T_{\alu b}{}^c&=&T_{ab}{}^c =0~, \qquad 
T_{\a\ad ,}{}_{\b}^j{}^\g_k= \hf \d_\a^\g  \,T_{\r\ad ,}{}_{\b}^j{}^\r_k~.
\label{constraints}
\eea
The gauge field $V_A$ also has to obey covariant constraints  to describe the vector multiplet. 
The vector multiplet constraints \cite{GSW} are 
\bea
F_\a^i{}_\b^j=-2\ve_{\a\b}\ve^{ij}\bar{W}~,\qquad 
F^\ad_i{}^\bd_j=2\ve^{\ad\bd}\ve_{ij}{W}~,\qquad
F_\a^i{}^\bd_j=0~.
\eea 

The solution to the constraints
is as follows:
\begin{subequations}
\bea
\{\cD_\a^i,\cD_\b^j\}&=&
4S^{ij}M_{\a\b}
+2\ve^{ij}\ve_{\a\b}Y^{\g\d}M_{\g\d}
+2\ve^{ij}\ve_{\a\b}\bar{W}^{\gd\dd}{\bar M}_{\gd\dd}
\non\\
&&
+2 \ve_{\a\b}\ve^{ij}S^{kl}J_{kl}
+4 Y_{\a\b}J^{ij}
-2\ve_{\a\b}\ve^{ij}\bar{W}{\mathbb Z}~,
\label{a-c1}\\
%%%%%%%%%%%%%%%%%%%%%%%%%%%%%%%%%%%%%%%%%%%%%
\{\cD_\a^i,\cDB^\bd_j\}&=&
-2\ri\d^i_j(\s^c)_\a{}^\bd\cD_c
+4 \big( \d^{i}_{j}G^{\d\bd}  +\ri G^{\d\bd}{}^{i}{}_{j} \big) M_{\a\d}
+4\big( \d^{i}_{j}G_{\a\gd}
+\ri G_{\a\gd}{}^{i}{}_{j}\big) {\bar M}^{\gd\bd} ~~~~~~
\non\\
&&
+8 G_\a{}^\bd J^{i}{}_{j}
-4\ri\d^i_jG_{\a}{}^\bd{}^{kl}J_{kl}
-2\big( \d^i_jG_{\a}{}^{\bd}
+\ri G_{\a}{}^{\bd}{}^i{}_j\big) {\mathbb J} ~, 
\label{a-c2}\\
{[}\cD_a,\cD_\b^j{]}&=& -\ri 
(\ts_a)^{\ad\g}\Big(
\d^j_kG_{\b\ad}
+ \ri G_{\b\ad}{}^{j}{}_k\Big)
\cD_\g^k
\non\\
&&
+{\frac\ri 2}\Big(({\s}_a)_{\b\gd}S^{jk}
-\ve^{jk}({\s}_a)_\b{}^{\dd}\bar{W}_{\dd\gd}
-\ve^{jk}({\s}_a)^{\a}{}_\gd Y_{\a\b}\Big)\cDB^\gd_k
\non\\
&&
+\hf R_a{}_\b^j{}^{cd}M_{{c}{d}}
+R_a{}_\b^j{}^{kl}J_{kl}
+\ri R_a{}_\b^j\,{\mathbb J}
+{\frac\ri 2}(\s_a)_\b{}^{\gd}\cDB_\gd^j\bar{W} {\mathbb Z}~.
\label{vector-spinor}
\eea
\end{subequations}
Here  the dimension-1 components of the torsion
obey the symmetry properties  
\bea
S^{ij}=S^{ji}~, \qquad Y_{\a\b}=Y_{\b\a}~, 
\qquad W_{\a\b}=W_{\b\a}~, \qquad G_{\a\ad}{}^{ij}=G_{\a\ad}{}^{ji}
\eea
and the reality conditions
\bea
\overline{S^{ij}} =  \bar{S}_{ij}~,\quad
\overline{W_{\a\b}} = \bar{W}_{\ad\bd}~,\quad
\overline{Y_{\a\b}} = \bar{Y}_{\ad\bd}~,\quad
\overline{G_{\b\ad}} = G_{\a\bd}~,\quad
\overline{G_{\b\ad}{}^{ij}} = ~G_{\a\bd}{}_{ij}.
\eea
The ${\rm U}(1)_R$ charges of the complex fields are:
\bea
{\mathbb J} \,S^{ij}=2S^{ij}~,\qquad
{\mathbb J}  \,Y_{\a\b}=2Y_{\a\b}~, \qquad
{\mathbb J} \, W_{\a\b}=-2W_{\a\b}~, \qquad
{\mathbb J}  \, W=-2W~.
\eea
The dimension-3/2 components of the curvature appearing in (\ref{vector-spinor}) 
have the following explicit form:
\begin{subequations}
\bea
R_a{}_\b^j{}_{cd}&=&
-\ri(\s_d)_{\b}{}^{\dd} T_{ac}{}_\dd^j
+\ri(\s_a)_{\b}{}^{\dd} T_{cd}{}_\dd^j
-\ri(\s_c)_{\b}{}^{\dd} T_{da}{}_\dd^j
~,
\label{3/2curvature-1}
\\
R_{\a\ad}{}_{\b}^j{}^{kl}
&=&
-{\ri}\ve^{j(k}\cDB_\ad^{l)}Y_{\a\b}
-{\ri}\ve_{\a\b}\ve^{j(k}\cDB^{\dd l)}\bar{W}_{\ad\dd}
-{\frac\ri 3}\ve_{\a\b}\ve^{j(k}\cDB_{\ad q}S^{l)q}
\non\\
&&
+{\frac43}\ve^{j(k}\cD_{(\a q}G_{\b)\ad}{}^{l)q}
+{\frac23}\ve_{\a\b}\ve^{j(k}\cD^\d_{q}G_{\d\ad}{}^{l)q}
~,
\\
R_{\a\ad}{}_{\b}^j&=&
-\cD_{\b}^jG_{\a\ad}
+{\frac\ri 3}\cD_{(\a k}G_{\b)\ad}{}^{jk}
+{\frac\ri 2}\ve_{\a\b}\cD^{\g}_kG_{\g\ad}{}^{jk}~.
\eea
\end{subequations}
The right-hand side of  (\ref{3/2curvature-1}) involves the dimension-3/2 components 
of the torsion which are expressed in terms of the dimension-1 tensors as follows: 
\begin{subequations}
\bea
&&T_{ab}{}_\gd^k\equiv(\s_{ab})^{\a\b}\cT_{\a\b}{}_{\gd}^{k}
-(\ts_{ab})^{\ad\bd}\cT_{\ad\bd}{}_{\gd}^{k}~,~~~
\\
&&\cT_{\a\b}{}_{\gd}^{k}
=
-{\frac14}\cDB_{\gd}^{k}Y_{\a\b}
+{\frac\ri 3}\cD_{(\a}^lG_{\b)\gd}{}^{k}{}_l
~,
\\
&&\cT_{\ad\bd}{}_{\gd}^{ k}
=
-{\frac14}\cDB_{\gd}^k\bar{W}_{\ad\bd}
-{\frac16}\ve_{\gd(\ad}\cDB_{\bd)l}S^{kl}
-{\frac\ri 3}\ve_{\gd(\ad}\cD^{\d}_qG_{\d\bd)}{}^{kq}
~.
\eea
\end{subequations}

The dimension-3/2 Bianchi identities are:
\begin{subequations}
\bea
\cD_{\a}^{(i}S^{jk)}&=&0~, \qquad 
\cDB_{\ad}^{(i}S^{jk)} = \ri\cD^{\b (i}G_{\b\ad}{}^{jk)}~,
\label{BI-3/2-1}
 \\
\cD_\a^i\bar{W}_{\bd\gd}&=&0~,\\
\cD_{(\a}^{i}Y_{\b\g)}&=&0~, \qquad 
\cD_{\a}^{i}S_{ij}+\cD^{\b}_{j}Y_{\b\a}=0~, \\
\cD_{(\a}^{(i}G_{\b)\bd}{}^{jk)}&=&0~, \\
\cD_\a^iG_{\b\bd}&=&
- \frac{1}{ 4}\cDB_\bd^iY_{\a\b}
+ \frac{1}{ 12}\ve_{\a\b}\cDB_{\bd j}S^{ij}
- \frac{1}{ 4}\ve_{\a\b}\cDB^{\gd i}\bar{W}_{\gd\bd}
- \frac{\ri }{ 3}\ve_{\a\b}\cD^{\g}_j G_{\g \bd}{}^{ij}~.
\eea
\end{subequations}
The Bianchi identities for the vector multiplet are
\begin{subequations}
\bea
\cDB^\ad_iW&=& 0~, 
\label{vector-Bianchi-chiral}\\ 
\Big( \frac{1}{ 4}\cD^{\g(i}\cD_\g^{j)}+S^{ij}\Big)W&=&
\Big( \frac{1}{ 4}\cDB_\gd^{(i}\cDB^{\gd j)}+\bar{S}^{ij}\Big)\bar{W} \equiv \S^{ij}~, 
\qquad \overline{\S^{ij}} = \S_{ij}~.
\label{vector-Bianchi}
\eea
\end{subequations}
Using the anti-commutation relations  (\ref{a-c1}) and (\ref{a-c2}),
the Bianchi identities (\ref{BI-3/2-1}) and (\ref{vector-Bianchi-chiral}),
one can check that eq. (\ref{vector-Bianchi}) implies
the following relations:
\bea
\cD_{\a}^{(i}\S^{jk)} = {\bar \cD}_{\ad}^{(i}\S^{jk)} =0~.
\eea
It should be pointed out that the vector multiplet field strength, 
$F_{ab}$,  is expressed in terms of
the covariantly chiral scalar  $W$ 
and its conjugate as follows:
\bea
F_{ab}&=&
- \frac{1}{8}(\s_{ab})_{\b\g}\cD^{\b k}\cD^{\g}_k{W}
- \frac{1}{ 8}(\ts_{ab})_{\bd\gd}\cDB^{\bd k}\cDB^{\gd}_k\bar{W}
\non\\
&&
-\frac{1}{ 4}\Big(
(Y_{ab}+W_{ab})(W+\bar{W})
+\frac{\ri }{ 2}\ve_{abcd}(Y^{cd}-W^{cd})(W-\bar{W})
\Big) ~.~~~~~~~~~
\eea

\subsection{Super-Weyl transformations}
The constraints (\ref{constraints}) were shown in  \cite{Howe} to be invariant 
under infinitesimal super-Weyl  transformations generated 
by a real  unconstrained parameter $U=\bar{U}$. 
We find the finite form of
such a 
transformation to be 
\begin{subequations}
\bea
\cD'{}_\a^i&=&\re^{U}\Big(\cD_\a^i+4(\cD^{\g i}U)M_{\g\a}-4(\cD_{\a k}U)J^{ki}
-(\cD_\a^iU) \,{\mathbb J} \Big)~,
\label{Finite_D}\\
\cDB'_{\ad i}&=&\re^{U}\Big(\cDB_{\ad i}+4(\cDB^{\gd}_{i}U)\bar{M}_{\gd\ad}
+4(\cDB_{\ad}^{k}U)J_{ki}
+(\cDB_{\ad i}U)\,{\mathbb J}\Big)~,
\label{Finite_Db} 
\\
\cD'_{\a\ad}
&=&\re^{2U}\Big(
\cD_{\a\ad}
+2\ri(\cDB_{\ad k}U)\cD_\a^k
+2\ri(\cD_\a^kU)\cDB_{\ad k}
+2(\cD^\g{}_\ad U)M_{\g\a}+2(\cD_\a{}^\gd U)\bar{M}_{\gd\ad}
\non\\
&&~~~~~
-4\ri(\cD^{\g k}U)(\cDB_{\ad k}U)M_{\g\a}
+4\ri(\cD_\a^{k}U)(\cDB^\gd_kU)\bar{M}_{\gd\ad}
\non\\
&&~~~~~
+8\ri(\cD_{\a}^{(k}U)(\cDB_\ad^{l)}U)J_{kl}
+{\frac\ri 2}(\cD_{\a}^{k}U)(\cDB_{\ad k}U)\,{\mathbb J}
\Big)
~.~~~~~~~~
\label{Finite_D_c}
\eea
\end{subequations}
These relations imply that the dimension-1 components of the torsion transform as
\begin{subequations}
\bea
W'_{\a\b}&=&\re^{2U}{W}_{\a \b}~,
\label{Finite_W}
\\
Y'_{\a\b}&=&\re^{2U}\Big(Y_{\a\b}
-(\cD^k_{(\a}\cD_{\b)k}U)
-4(\cD^k_{(\a}U)(\cD_{\b)k}U)\Big)
\label{Finite_Y}~,\\
S'_{ij}&=&\re^{2U}\Big(S_{ij}
-(\cD^\g_{(i}\cD_{\g j)}U)
+4(\cD^\g_{(i}U)(\cD_{\g j)}U)\Big)
\label{Finite_S}~,
\\
G'_{\a\ad}&=&
\re^{2U}\Big(G_{\a\ad}
-{\frac14}[\cD_\a^k,\cDB_{\ad k}]U
-2(\cD_\a^kU)(\cDB_{\ad k}U)
\Big)
~,
\label{Finite_G}
\\
G'_{\a\ad}{}^{ij}&=&\re^{2U}\Big(G_{\a\ad}{}^{ij}
+{\frac\ri 2}[\cD_\a^{(i},\cDB_\ad^{j)}]U\Big)
~.
\label{Finite_Gij}
\eea
\end{subequations}
In the infinitesimal case, the above transformation laws reduce 
to those given in \cite{Howe}.
Of special importance for our consideration below is the fact 
that the right-hand side in (\ref{Finite_Gij}) contains no contribution 
quadratic in derivatives of $U$.

The super-Weyl transformation of the vector multiplet field strength is 
\bea
W'=\re^{2U}W~.
\label{super-Weyl-W}
\eea 
Using this result, one can derive 
the super-Weyl transformation of the descendant $\S^{ij}$ 
introduced in (\ref{vector-Bianchi}). It is 
\bea
\S'_{ij} = {\rm e}^{4U} \S_{ij}~.
\label{super-Weyl-S}
\eea

\subsection{Partial gauge fixing I}
The torsion $G_{\a\ad}{}^{ij}$ turns out to be a pure gauge degree of 
freedom with respect to the super-Weyl symmetry.
This means that 
\bea
G_{\a\ad}{}^{ij}
=-{\frac\ri 2}[\cD_\a^{(i},\cDB_\ad^{j)}] {\bm U}~,
\label{G}
\eea
for some real scalar superfield $\bm U$.
The simplest way to see this is to follow Howe's procedure
of introducing the minimal supergravity  multiplet \cite{Howe}.

Suppose that the Abelian vector multiplet, which was introduced 
in subsection \ref{subsec3.1},  is such that $W\neq 0$ at each point of the superspace.
Under the super-Weyl and local ${\rm U}(1)_R$ transformations, 
the field strength changes as 
\bea
W~ \to  ~\re^{2(U -{\rm i} L)}W~.
\eea
Such a  combined transformation acts on $G_{\a\ad}{}^{ij}$ according to eq.
(\ref{Finite_Gij}), for $G_{\a\ad}{}^{ij}$ is neutral with respect to $\mathbb J$.
Since the transformation parameters $U$ and $L$ are real and unconstrained, 
it is in our power to choose the gauge
\be
W=1~
\label{W=1}
\ee
which completely fixes the super-Weyl and local ${\rm U}(1)_R$ symmetries.
What are the implications of this gauge fixing? First of all, 
the condition that $W$ is covariantly chiral implies that 
$0=\cDB^\ad_i W=-2\ri \,\Phi^\ad_i=0$,
and therefore
\bea
\Phi_\a^i=\Phi^\ad_i=0~.
\label{U(1)-gauge}
\eea
Since  the  spinor ${\rm U}(1)_R$ connections vanish, the gauge condition (\ref{W=1})
and the Bianchi identity (\ref{vector-Bianchi}) lead to 
\bea
S^{ij}=\bar{S}^{ij}~.
\eea
Similar arguments give
\bea
0&=&\cD_\a^i\cDB_\bd^j\bar{W}
=2\ri\ve^{ij}(\s^a)_{\a\bd}\cD_a\bar{W}+4\ve^{ij}G_{\a\bd}\bar{W}-4\ri G_{\a\bd}{}^{ij}\bar{W}
\non\\
&=&-4\ve^{ij}\F_{\a\bd}+4\ve^{ij}G_{\a\bd}-4\ri G_{\a\bd}{}^{ij}
\non
\eea
and therefore
\be
G_{\a\bd}{}^{ij}=0~, \qquad
\F_{\a\bd}=G_{\a\bd}~.
\ee
The first equation here tells us that $G_{\a\bd}{}^{ij}$ vanishes upon imposing the 
 super-Weyl + local ${\rm U}(1)_R$ gauge condition (\ref{W=1}).
Recalling the super-Weyl transformation law of $G_{\a\bd}{}^{ij}$,
eq. (\ref{Finite_Gij}), we conclude that the general form for $G_{\a\bd}{}^{ij}$
is given by eq. (\ref{G}).

\subsection{Partial gauge fixing II}

In the above consideration, the vector multiplet played the role 
of a useful technical tool that allowed us to prove eq. (\ref{G}). 
Since eq. (\ref{G}) has been justified, we can undo the gauge condition 
(\ref{W=1}) and return to the general case. 
Due to  (\ref{G}) and the super-Weyl transformation (\ref{Finite_Gij}), 
we can  use the  super-Weyl gauge freedom to choose
\bea
G_{\a\bd}{}^{ij}=0~.
\label{G2}
\eea
In this gauge, let us introduce new covariant derivatives $\tilde{\cD}_A$ defined 
by the rule:
\bea
\tilde{\cD}_\a^i = \cD_\a^i~, \qquad
\tilde{\cD}_a=\cD_a-\ri \,G_a \,{\mathbb J}~.
\eea
Then, making use of  the (anti) commutation relations (\ref{a-c1}), (\ref{a-c2}) 
and (\ref{vector-spinor}), one can readily check the covariant derivatives $\tilde{\cD}_A$ 
have no ${\mathbb J}$-curvature, 
$\tilde{R}_{AB}=0$,  and therefore the corresponding connection 
$\tilde{\F}_A$ is flat. We can  choose $\tilde{\F}_A=0$ by
applying an appropriate local ${\rm U}(1)_R$ transformation.
As a result, the superspace geometry proves to reduce to the one used
in \cite{KLRT_M} for the description of general supergravity-matter systems.
This geometry corresponds to Grimm's curved superspace setting \cite{Grimm}.

Let us suppose that we have chosen the super-Weyl gauge condition (\ref{G2})
and also fixed the local ${\rm U}(1)_R$ symmetry by setting $\F_\a^i =0$.
Eq. (\ref{G2}) does not completely fix  the super-Weyl symmetry. 
In accordance with (\ref{Finite_Gij}), the residual gauge freedom is described 
by a parameter $U$ constrained as 
\be
 [\cD_\a^{(i},\cDB_\ad^{j)}]U=0~.
 \label{Ucon}
 \ee
As pointed out in \cite{KLRT_M}, the general solution of this equation is 
\bea
U = \frac{1}{4} (\s+\sba)~, \qquad {\bar \cD}^\ad_i \s =0~,
\qquad {\mathbb J}\, \s =0~.
\eea
Here the parameter $\s$ is covariantly chiral but otherwise 
arbitrary.
As follows from (\ref{Finite_D}) and (\ref{Finite_Db}), 
such a super-Weyl transformation must be accompanied 
by the following compensating ${\rm U}(1)_R$-transformation 
\be
\cD'_A=\re^{\ri L {\mathbb J}}\, \cD_A\, \re^{-\ri L {\mathbb J}}~, 
\qquad L=\frac{\ri}{4}(\s - \sba)
\ee
to preserve the gauge condition $\F_\a^i =0$.
The resulting transformation is 
\begin{subequations}
\bea
\cD'{}_\a^i&=&\re^{\hf\sba}\Big(\cD_\a^i+(\cD^{\g i}\s)M_{\g\a}-(\cD_{\a k}\s)J^{ki}\Big)~,
\label{Finite_D2}\\
\cDB'_{\ad i}&=&\re^{\hf\s}\Big(\cDB_{\ad i}+(\cDB^{\gd}_{i}\sba)\bar{M}_{\gd\ad}
+(\cDB_{\ad}^{k}\sba)J_{ki}\Big)~.
\label{Finite_Db2} 
\eea
\end{subequations}
In the infinitesimal case, this super-Weyl transformation reduces to that given in \cite{KLRT_M}.
The finite super-Weyl transformations of the covariant derivatives, 
eqs. (\ref{Finite_D2}) and (\ref{Finite_Db2}), 
and of  various components of the torsion
were given in \cite{KT-M-field}.

It is interesting to point out analogies between the 4D $\cN=2$ superspace formulation 
considered with that for 5D $\cN=1$ conformal supergravity\footnote{The superconformal tensor 
calculus in five dimensions was developed in \cite{Ohashi,Bergshoeff}.}
 \cite{KTMfiveWeyl}.
In the five-dimensional case, 
the super-Weyl transformations 
are also generated by a real unconstrained  parameter \cite{KTMfiveWeyl}.
Moreover, the corresponding superspace torsion includes 
a vector-isovector component $C_{\hat a}{}^{ij} = C_{\hat a}{}^{ji}$, 
with the lower index being 5D vector,  
which can be gauged away by the super-Weyl transformations. 
This superfield is the 5D analogue of $G_{\a \ad}{}^{ij}$.
In the gauge $C_{\hat a}{}^{ij} = 0$, the super-Weyl parameter 
obeys a constraint which is similar to  (\ref{Ucon}).

\section{Curved projective superspace}
\setcounter{equation}{0}
Matter couplings in supergravity are described in \cite{KLRT_M} in terms
of covariant projective supermultiplets. In this section, we first generalize the concept 
of covariant projective supermultiplets to the case of Howe's formulation for 
conformal supergravity, and then we present a locally supersymmetric 
and super-Weyl invariant action.

\subsection{Covariant $O(n)$ supermultiplets}
Consider a completely symmetric isotensor superfield
$F^{i_1 \dots i_n} = F^{(i_1 \dots i_n)}$.
For simplicity, we assume it  to be
neutral with respect to the central charge generator $\mathbb Z$ in 
(\ref{CovDev}),  ${\mathbb Z}\,  F^{i_1 \dots i_n} = 0$, although this condition 
is not necessary for the discussion below.
We require $F^{i_1 \dots i_n}$ to obey the constraints\footnote{Constraints 
of the form (\ref{F-constraints}) have a long history in rigid $\cN=2$ supersymmetry.
${}$For $n=1$ they define an on-shell hypermultiplet \cite{Sohnius}; 
the supermultiplet becomes off-shell if one allows for a non-vanishing intrinsic 
central charge, ${\mathbb Z} F^i \neq0$. The case $n=2$ was considered in 
\cite{BS,SSW,KLR} and corresponds to the off-shell $\cN=2$ tensor multiplet \cite{Wess}
provided $F^{ij}$ is real.
The case $n=4$ was briefly discussed in \cite{SSW} in the context of superactions, and 
it also played a key role in the relaxed hypermultiplet construction \cite{HST}.
The constraints for arbitrary  $n>2$ first appeared in \cite{HSW}. 
These constraints were shown in \cite{KLT,LR} to provide
alternative off-shell formulations for the hypermultiplet if $n=2m$, with $m=2,3 \dots$, 
and $F^{i_1 \dots i_{2m} }$ is chosen to be real.}
\bea
 \cD^{(j}_\a F^{i_1 \cdots i_{n})}={\bar \cD}^{(j}_\ad F^{i_1 \cdots i_{n})}=0~.
 \label{F-constraints}
\eea
Using the anti-commutation relations (\ref{a-c1}) and (\ref{a-c2}), 
one can check that these constraints are consistent 
provided the following conditions hold:\\
(i) $F^{i_1 \dots i_n} $ is neutral with respect to $\mathbb J$, 
\be
{\mathbb J}\,  F^{i_1 \dots i_n} = 0~;
\label{no-charge}
\ee
(ii)
$F^{i_1 \dots i_n} $ 
is scalar with respect to the Lorentz group, 
\be 
M_{ab} F^{i_1 \dots i_n} =0~.
\label{no-Lorentz-indices}
\ee
Thus, the transformation law (\ref{tensor-K}) in the case of  $F^{i_1 \dots i_n} $ becomes
\bea
\d_\cK F^{i_1 \dots i_n} = \big( K^C \cD_C + K^{kl} J_{kl} \big)F^{i_1 \dots i_n}~, 
 \qquad K^{kl} J_{kl}  \,F^{i_1\cdots i_n} 
= \sum_{l=1}^{n} K^{i_l}{}_j \,F^{j i_1\cdots  \widehat{i_l} \cdots i_n} ~.~~~
\label{F-var-1} 
\eea
One can associate with  $F^{i_1 \dots i_n} $ a holomorphic tensor field on ${\mathbb C}P^1$, 
$F^{(n)}(u^+)$, defined as 
\bea
F^{(n)}(u^+)= u^+_{i_1}\cdots u^+_{i_{n}}\,F^{i_1\cdots i_{n}}~, 
\qquad F^{(n)}(c\,u^+)=c^n \,F^{(n)}(u^+)~, \quad c\in \mathbb{C} \setminus \{0\}~,~~
\label{F-n}
\eea
with $u^+_i \in {\mathbb C}^2 \setminus \{0\}$ homogeneous 
coordinates for ${\mathbb C}P^1$.

It is useful to take the auxiliary variables $u^+_i$ 
to be inert\footnote{This is similar to the
approach often used  in the context  of higher spin field theories, see e.g. \cite{FMS}.}
under the local ${\rm SU}(2)_R$ group, 
that is $[J_{kl} , u^+_i ]=0$, 
for their  sole role is to describe  $F^{i_1\cdots i_{n}}$
in terms of  the index-free object $ F^{(n)}(u^+)$.
Then, the transformation law (\ref{F-var-1}) can be rewritten as
\bea
\d_\cK F^{(n)} 
&=& \Big( K^{{C}} \cD_{{C}} + K^{kl} J_{kl} \Big)F^{(n)} ~,  
\non \\ 
K^{kl} J_{kl}  F^{(n)}&=& -\frac{1}{(u^+u^-)} \Big(K^{++} D^{(-1,1)} 
-n \, K^{+-}\Big) F^{(n)} ~, \qquad 
K^{\pm \pm } =K^{ij}\, u^{\pm}_i u^{\pm}_j ~,
\label{F-var-2}   
\eea 
where
\bea
D^{(-1,1)}
:=u^{-i}\frac{\partial}{\partial u^{+ i}} ~.
\label{5}
\eea
Eq. (\ref{F-var-2}) involves an additional complex two-vector,  $u^-_i$, 
which has  to be linearly independent of $u^+_i$,  
that is $(u^+u^-) := u^{+i}u^-_i \neq 0$, and is otherwise completely arbitrary.
It is important to note that since the $u^+_i$
are fixed and  constant, $F^{(n)} (u^+)$ is not isoscalar. 
In this approach, the $u^+_i$ serve merely to 
totally symmetrize all ${\rm SU}(2)_R$ indicies. 

Without imposing the constraints (\ref{F-constraints}) and their corollaries 
(\ref{no-charge}) and (\ref{no-Lorentz-indices}), 
the above consideration can be naturally generalized. Namely, one can allow 
$F^{i_1 \dots i_n} = F^{(i_1 \dots i_n)}$ to carry any number of Lorentz indices and
have a non-vanishing $\mathbb J$-charge. Let  $ F^{(n)}(u^+)$ be the homogeneous 
polynomial of degree  $n$ associated with $F^{i_1 \dots i_n} $.
An operation of  multiplication is naturally defined in the space of  such polynomials, 
for  given two homogeneous polynomials  
$F^{(n)}(u^+)$ and $F^{(m)}(u^+)$, 
their product $F^{(n+m)}(u^+):=F^{(n)}(u^+)\,F^{(m)}(u^+)$
is a homogeneous polynomials of degree $(n+m)$. 
If one introduces the differential operators 
$\cD^+_{ \a}:=u^+_i\,\cD^i_{ \a}$ and 
${\bar \cD}^+_{\dot  \a}:=u^+_i\,{\bar \cD}^i_{\dot \a}$, 
then 
\bea
\cD^+_{ \a} F^{(n)} (u^+)
= 
u^+_j u^+_{i_1}\cdots u^+_{i_{n}}\,\cD^{(j}_\a F^{i_1 \cdots i_{n})} ~, \qquad
{\bar \cD}^+_{ \ad} F^{(n)} (u^+)= 
u^+_j u^+_{i_1}\cdots u^+_{i_{n}}\,{\bar \cD}^{(j}_\ad F^{i_1 \cdots i_{n})} ~
\non
\eea
are homogeneous polynomials of degree $(n+1)$.
Here we have used the fact  that the auxiliary variables $u^+_i$ 
are inert under the local ${\rm SU}(2)_R$ group, $[J_{kl} , u^+_i ]=0$. 

The example of $F^{(n)}$'s considered
can naturally be extended to define more general 
{\it isotwistor} superfields.  
They are introduced similarly to 
the consideration given in the appendix in \cite{KLRT_M}. 
The only difference from \cite{KLRT_M}
is that now an isotwistor superfield may have 
a non-vanishing  $\mathbb J$-charge.

Let us now return to the constraints (\ref{F-constraints}). 
They are  equivalent to 
\bea
\cD^+_{ \a} F^{(n)} = {\bar \cD}^+_{\dot  \a} F^{(n)} =0~.
\label{F-constraints-2}
\eea
When acting on isotwistor superfields,
the differential operators $ \cD^+_{ \a} $ and ${\bar \cD}^+_{\dot  \a}$
obey the following anti-commutation relations: 
\begin{subequations}
\bea
\{  \cD^+_{ \a} , \cD^+_{ \b} \}
&=&
4 S^{++}M_{\a \b}
+4Y_{\a \b}J^{++}~, 
\label{analyt0}
\\ 
\{\cD_\a^+,\cDB_\bd^+\}&=&
4\ri G^\g{}_\bd{}^{++}M_{\a\g}
-4\ri G_{\a}{}^{\gd}{}^{++}\bar{M}_{\bd\gd}
+8 G_{\a \bd} J^{++}
-2\ri \,G_{\a\bd}{}^{++}{\mathbb J}~,
\label{analyt}
\eea
\end{subequations}
where we have defined 
\be
J^{++}:= u^+_iu^+_j J^{ij}~, \qquad S^{++}:= u^+_iu^+_j S^{ij}~,
\ee
and similarly for $G_{\a\bd}{}^{++}$.
The constraints (\ref{F-constraints-2}) are consistent 
because the integrability condition $J^{++}  F^{(n)} =0$ holds identically.
The other integrability conditions for the constraints (\ref{F-constraints-2}) 
are: ${\mathbb J} \,F^{(n)} =0$ and $M_{ab} F^{(n)} =0$.
${}$Following \cite{KLRT_M}, the superfield $F^{(n)}$ 
will be  called a covariant $O(n)$ supermultiplet.

As an example  of $O(n)$  supermultiplets, we can consider the $O(2)$ 
multiplet 
\be
\S^{++} = u^+_i  u^+_j \S^{ij}~, 
\ee
with $\S^{ij}$ defined in (\ref{vector-Bianchi}).

Using $O$-type supermultiplets, $F^{(n)}$ and  $H^{(m)}$, 
one can construct  covariant rational supermultiplets 
of the form 
\bea
R^{(n-m)}(u^+) = 
\frac{F^{(n)}(u^+)} {H^{(m)}(u^+)}~, 
\eea
which correspond to meromorphic tensor fields on ${\mathbb C}P^1$.
The $R^{(p)}(u^+) $ possesses properties which are completely similar to  (\ref{F-var-2}) and
(\ref{F-constraints-2}). 
In the rigid supersymmetric case, rational supermultiplets were introduced in \cite{LR}.
The above superfields  are examples of covariant 
projective supermultiplets we will now introduce.  

\subsection{Covariant projective supermultiplets}

By definition, 
a covariant projective supermultiplet of weight $n$,
$Q^{(n)}(z,u^+)$, is a {\it scalar} superfield that lives on  $\cM^{4|8}$, 
is holomorphic on an open domain of ${\mathbb C}^2 \setminus  \{0\}$ 
 with respect to 
the homogeneous coordinates $u^{+}_i $  for ${\mathbb C}P^1$,
and is characterized by the  conditions:\\
(i) it obeys the covariant  constraints 
\be
\cD^+_{\a} Q^{(n)}  = {\bar \cD}^+_{\ad} Q^{(n)}  =0~;
\label{ana}
\ee  
(ii) it is  a homogeneous function of $u^+$ 
of degree $n$, that is,  
\be
Q^{(n)}(z,c\,u^+)\,=\,c^n\,Q^{(n)}(z,u^+)~, \qquad c\in \mathbb{C}\setminus \{ 0 \}~;
\label{weight}
\ee
(iii) it is neutral  with respect to  $\mathbb J$:
\bea
{\mathbb J} \,Q^{(n)}(z,u^+)=0
\eea
(iv)  the supergravity gauge transformations act on $Q^{(n)}$ 
as follows:
\bea
\d_\cK Q^{(n)} 
&=& \Big( K^{{C}} \cD_{{C}} + K^{kl} J_{kl} \Big)Q^{(n)} ~,  
\non \\ 
K^{kl} J_{kl}  Q^{(n)}&=& -\frac{1}{(u^+u^-)} \Big(K^{++} D^{(-1,1)} 
-n \, K^{+-}\Big) Q^{(n)} ~. 
\label{harmult1}   
\eea 
Using eqs.~(\ref{analyt0}) and (\ref{analyt}) one can see that these definitions
are consistent. The integrability condition for the constraints (\ref{ana})  
is  $J^{++}  Q^{(n)} =0$,  and clearly it holds identically. 

What are admissible  super-Weyl transformations  
of projective supermultiplets?
Assuming that $Q^{(n)}$  transforms homogeneously under the super-Weyl transformations,
the constraints (\ref{ana}) uniquely fix its transformation law:
\bea
\d_U Q^{(n)}=2nU\, Q^{(n)}~.
\eea

${}$On the space of covariant projective supermultiplets, 
one can introduce a generalized (smile) conjugation
$Q^{(n)} (u^+) \to \widetilde{Q}^{(n)} (u^+)$, with $\widetilde{Q}^{(n)} $ 
also being a covariant projective supermultiplet. 
The smile-conjugation is defined in \cite{KLRT_M}.
If $n$ is even, one can consistently define real supermultiplets.

If one partially fixes the super-Weyl symmetry as in (\ref{G2}) 
as well as  imposes the ${\rm U}(1)_R$ gauge condition (\ref{U(1)-gauge}), the above definitions 
and properties reduce to those given in \cite{KLRT_M}.

\subsection{Action principle}
Within the curved superspace setting under consideration, 
the  construction of supersymmetric action principle is practically identical to that 
given in \cite{KLRT_M}. 
Let  $\cL^{++}$ be a real projective multiplet of weight two, 
with the  super-Weyl transformation law
\be
\d_U \cL^{++}= 4U \, \cL^{++} ~.
\label{L(++)super-Weyl}
\ee 
Associated with $\cL^{++}$ is the following functional: 
\bea
S&=&
\frac{1}{2\pi} \oint (u^+ \rd u^{+})
\int \rd^4 x \,{\rm d}^4\q {\rm d}^4{\bar \q}\,E\, \frac{W{\bar W}\cL^{++}}{(\S^{++})^2}~, 
\qquad E^{-1}= {\rm Ber}(E_A{}^M)~.
\label{InvarAc}
\eea
By construction, this functional is  invariant under  re-scalings
$u_i^+(t)  \to c(t) \,u^+_i(t) $, for an arbitrary function
$ c(t) \in {\mathbb C}\setminus  \{0\}$, 
where $t$ denotes the evolution parameter 
along the closed integration contour.
Since ${\mathbb J}\,E=0$ and ${\mathbb J}\,(W\bar W ) =0$, 
$S$ is invariant under the local U(1) transformations.
Using this observation, the above functional  
can be shown to be invariant under arbitrary supergravity gauge 
transformations, eqs. (\ref{tau}) and (\ref{tensor-K}), 
in complete analogy with  \cite{KLRT_M}. 
Since $E$ is invariant under the super-Weyl transformations,
\be
\d_{U} E=0~,
\ee
the transformation laws (\ref{super-Weyl-W}), (\ref{super-Weyl-S}) and 
(\ref{L(++)super-Weyl}) tell us that 
$S$ is  super-Weyl invariant.

In the super-Weyl and local ${\rm U}(1)_R$ gauge defined by eqs.  (\ref{G2})  and (\ref{U(1)-gauge}), 
the action (\ref{InvarAc})  reduces to that proposed in \cite{KLRT_M}.

The locally supersymmetric and super-Weyl invariant action (\ref{InvarAc}) 
is suitable to describe the dynamics of general $\cN=2$ supergravity-matter system
including the formulations of Poincar\'e supergravity introduced in \cite{KLRT_M,K_2008}. 
In particular this is  true for chiral actions of the form
\bea
S_{\rm c}= \int \rd^4 x \,{\rm d}^4\q \, \cE \, \cL_{\rm c} &+& {\rm c.c.}~,
\qquad {\bar \cD}_\ad \cL_{\rm c} =0 ~, \quad {\mathbb J} \, \cL_{\rm c} = 
-4 \cL_{\rm c}~, \quad \d_U \cL_{\rm c}= 4U  \cL_{\rm c} ~,~~~
\eea
with $\cE$ the chiral density \cite{Muller,KT-M-normal}.
The latter follows from the fact that $S_{\rm c}$
admits the following representation \cite{K_2008}:
\bea
S_{\rm c}&=&\frac{1}{2\pi} \oint (u^+ \rd u^{+})
\int \rd^4 x \,{\rm d}^4\q {\rm d}^4{\bar \q}\, E\,
\frac{{ W}{\bar { W}}  \cL^{++}_{\rm c} }{({ \S}^{++})^2 }~, \non \\
\cL^{++}_{\rm c} &=&
 -\frac{1}{4} { V} \,\Big\{ \Big( (\cD^{+})^2+4{S}^{++}\Big) \frac{\cL_{\rm c}}{ W}
+\Big( (\cDB^{+})^2+4\bar{S}^{++}\Big)
\frac{{\bar \cL}_{\rm c} }{\bar { W}} \Big\}~,~~~~~~
\eea
with $V(u^+)$  the tropical prepotential for the vector multiplet 
with field strength $ W$, see \cite{KLRT_M}
for the definition of $V(u^+)$.

\section{Conclusion}
\setcounter{equation}{0}
${}$For many years, Howe's superspace formulation for $\cN=2$ conformal 
supergravity \cite{Howe} has remained a nice theoretical construction of purely academic 
interest. In the present paper, we demonstrated that  the curved superspace setting of 
\cite{Howe} is ideally suited for the construction 
of various matter couplings as well as a superspace action. For practical calculations, 
however, it is useful to work in the super-Weyl and local ${\rm U}(1)_R$ gauge 
 (\ref{G2})  and (\ref{U(1)-gauge}), in which the general supergravity-matter 
 systems reduce to those presented in \cite{KLRT_M}.\\

\noindent
{\bf Acknowledgements:}\\
We acknowledge the hospitality of  the 2008 Simons
Workshop in Mathematics and Physics where this project was initiated.
The work of SMK and GT-M was supported  in part by the Australian Research Council.
The research of MR  was supported in part by NSF grant no.~PHY-06-53342.
After 15 November 2008, GT-M is supported by the endowment 
 of the John S.~Toll Professorship, the University of 
 Maryland Center for String \& Particle Theory, and
 National Science Foundation Grant PHY-0354401.

\small{

}

\end{document}